\documentclass[a4paper,11pt]{article}
\usepackage[english]{babel}
\usepackage[T1]{fontenc}
\usepackage{textcomp}
\usepackage{epsfig}
\usepackage{latexsym}
\usepackage{graphicx}
\usepackage{amsfonts,amssymb,bm}
\usepackage{color}
\newcounter{llista}

\newtheorem{theorem}{Theorem}
\newtheorem{lemma}{Lemma}
\newtheorem{proposition}{Proposition}

\newtheorem{definition}{Definition}

\begin{document}
\author{J Llosa$^1$\thanks{e-mail address: pitu.llosa@ub.edu}\, A Molina$^1$\thanks{e-mail address: alfred.molina@ub.edu}  and D Soler$^2$\thanks{e-mail address: dsoler@eps.mondragon.edu}\\
\small $^1$ Dept. F\'{\i}sica Fonamental, Universitat de Barcelona, Spain\\
\small $^2$ Mechanical and Industrial Production Dept., Mondragon Unibertsitatea, Spain} 
      
\title{A relativistic generalisation of rigid motions}


\maketitle

\begin{abstract}
Radar-holonomic  congruences of wordlines are proposed as a weaker substitute for the too restrictive class of Born-rigid motions. The definition is expressed as a set of differential equations. Integrability conditions and Cauchy data are
studied. We finally obtain an example of a radar-holonomic congruence containing a given worldline with a given value of the rotation on this line.
\end{abstract}

\medskip\noindent
Keywords: rigid motions, congruences of worldlines, strain rate tensor,
radar metric \\
PACS number: 0420C \\ 
MSC number: 53B20, 83C99

\section{Introduction \label{S0}}
Although in General Relativity physical laws are expressed independently
of any specific reference frame, the design of an experiment and the
subsequent analysis of its results are often associated to a most suitable
reference frame, which embodies the ``laboratory frame''.

Often is tacitly assumed that this laboratory frame is rigid although, for several reasons, no consistent relativistic extension of the notion of rigidity has been defined yet.
Indeed, although Born's definition \cite{BORN09} of relativistic rigidity seems the most natural relativistic extension of the Newtonian notion, it has the drawback of being inconsistent in most cases. As proved by Herglotz and Noether \cite{HER-NOE10}, even in the most symmetric case of Minkowski spacetime, the only allowed Born-rigid motions are accelerated but rotationless motions or rotating motions with constant angular velocity around an origin in uniform motion (which are actually Killing motions).

This theoretical lack is obviated in most experimental dessigns. Think for instance in the kind of experiments using a resonant cavity to measure anisotropies in the speed of light \cite{BRILL}. A Fabri-Perot cavity is used a standard of length and it is tacitly assumed that this cavity is rigid; and consequently the cavity is made of a material ``as rigid as possible'' ---whatsoever that could mean and despite that this notion is not consistently defined in relativity.

This difficulty to find a theoretically consistent characterization of the laboratory reference frame is genuinely relativistic. In Newtonian mechanics there is no trouble in appealing to rigid reference frames, which are embodied by ideal rigid bodies or their imaginary prolongations \cite{EINS} (although ideal rigid bodies do not actually exist, even in Newtonian physics). 

At this point, and for the sake of clarity, it is convenient to distinguish the notions of (ideal) {\em rigid body} and {\em rigid motion}. In a {\em rigid motion} the {\em distance} between any couple of points remains constant along the motion, no matter how this configuration is maintained. Think, for instance, of a swarm of spaceships such that the travel plan of each one has been arranged so that it keeps at a constant distance of any other. The notion of rigid body needs one more ingredient: it always follows a rigid motion due to the internal constraint forces mutually exerted by the different parts of the body. 

In Newtonian mechanics none of these notions has inconsistences, appart form the fact that an ideal rigid body implies infinitely big elasticity moduli. Instead, in relativistic mechanics the very notion of rigid body is inconsistent because those internal constraint forces would imply signals propagating at an infinite speed. However, the notion of rigid motion is of a kinematical nature and it does not imply anything about what is done to keep it.

Even though, no consistent and satisfactory definition of relativistic {\it rigid motion} has been set up yet. What shall we take as the ``distance'' which is preserved along the motion? If one takes the radar distance, which seems the natural candidate, as it is done in Born's proposal \cite{BORN09}, then we have Herglotz-Noether negative result. 

A common feature of Newtonian rigid motions is that each one is unambiguously determined by the giving of the trajectory of one point together with the motion's vorticity along that line (that is, the angular velocity). The relativistic extension of this idea suggests that \cite{AUD} generalised Fermi coordinates are the natural coordinates to describe an accelerated rotating reference frame, and the congruence of world lines at rest relatively to this frame follows a rigid motion.

Generalised Fermi coordinates \cite{NIZ} are defined on the basis of an origin worldline $\gamma_O$ and three spatial axis that are Fermi-Walker transported along $\gamma_O$, with an arbitrary ascribed rotational motion. So that, the worldline of any point $P$ in the laboratory in these Fermi coordinates is: $\gamma_P\equiv\{(\tau,X^1,X^2,X^3)\}$, with $X^i=$constant, $i=1,2,3$. One would then be tempted to say that any place $P$ in the laboratory is at rest in the Fermi reference frame or, shortly, {\it $P$ is at rest relatively to $O$.}

The above reasoning is however unsatisfactory because this notion of rest is not transitive \cite{BEL94}, namely, if  $P$ is at rest relatively to $O$ and $Q$ is at rest relatively to $P$ do not imply necessarily that $Q$ is at rest with respect to $O$.

The relativistic generalization of rigid motions needs to be formulated in terms of a spacetime manifold (${\cal V}_4$, $g$). A {\it motion }is then defined by a 3-parameter congruence of timelike worldlines, $\mathcal{E}_3$, $x^\alpha (t)=\varphi ^\alpha (t,\,y^1,\,y^2,\,y^3)$ where $\,y^1,\,y^2,\,y^3$ are the parameters. In its turn, the congruence  is determined by its unit timelike velocity field $\mathbf{u}$
\begin{equation}  \label{E.2a}
u^\alpha (x)\,,\quad \mbox{ with} \qquad g_{\mu \nu }u^\mu u^\nu =-1
\end{equation}

The rest space of the congruence is the quotient space, where cosets are worldlines in $\mathcal{E}_3$, and will be denoted by the same symbol\footnote{The kinematics of a timelike congruence in spacetime has been studied in refs.  \cite{Zelmanov}, \cite{Cattaneo} and \cite{Ehlers93}, among others}. The radar metric 
\begin{equation}  \label{E.3}
{\hat g_{\alpha \beta }:=g_{\alpha \beta }+u_{\alpha}u_{\beta}}
\end{equation}
is associated to the infinitesimal radar distance $d\widehat{l}_R^2:=\widehat{g}_{\mu \nu }(x)dx^\mu dx^\nu$
between two neighbouring worldlines. This quantity does not define a distance on $\mathcal{E}_3$ because it
is not usually constant along the motion. Only in case that the Born-rigidity condition \cite{BORN09} holds, 
\begin{equation}  \label{E.4}
\Sigma_{\alpha\beta} := \mathcal{L}_\mathbf{u}\,\hat g_{\alpha\beta} = 0  \,,
\end{equation}
$\hat{g}_{\alpha \beta }$ defines a Riemannian metric on $\mathcal{E}_3$.

The above condition (\ref{E.4}) consists of six independent first order partial differential equations with three independent unknowns, namely $u^i$, $i=1,2,3$, just like in the Newtonian case.

The class of Born-rigid motions would generalize Newtonian rigid motions also because some spatial distance between points in space is conserved. Unfortunately, the Herglotz-Noether theorem \cite{HER-NOE10} states that,
even in Minkowski spacetime, the class of Born-rigid motions is narrower than sought and motions combining arbitrary acceleration and rotation are not encompassed by this class.

However, this shortness should not be surprising. Indeed, six first order partial differential equations for only three unknown functions unavoidably entail integrability conditions, which yield additional equations. In their turn, these will lead to new integrability conditions and so on. The process of completing the partial differential system (\ref{E.4}) ends up with a set of equations which is too restrictive for our desideratum, namely, six degrees of freedom: three for the motion of the origin and three for the angular velocity.

In a recent work \cite{LLOSA97} by one of us, 2-parameter congruences in a (2+1)-dimensional spacetime, $\mathcal{V}_3$, were considered as a simplification where the condition (\ref{E.4}) is still too restrictive
---three partial differential equations for two unknowns: $u^1(x)$ and $u^2(x)$. Then, the vanishing of shear $\displaystyle{\sigma_{\alpha\beta} := \Sigma_{\alpha\beta} - \frac12 \Sigma^\mu_\mu \, \hat g_{\alpha\beta} }$, $\alpha,\beta = 0,1,2 $ was advanced as a candidate to substitute the condition of Born-rigidity.

In a (2+1) dimensional spacetime, this condition (which is equivalent to conformal rigidity) reads: 
\begin{equation}  \label{E.5a}
\sigma_{\alpha \beta }=0\qquad \alpha ,\beta =0,1,2
\end{equation}
and yields two independent partial differential equations. 
Since the number of unknown functions is also two, the existence of congruences fulfilling condition (\ref{E.5a}) can be studied by standard methods in partial differential systems: given a non-characteristic surface $\mathcal{S}_2\subset \mathcal{V}_3$ and Cauchy data on it, a unique analytic solution of (\ref{E.5a}) in a neighbourhood of $\mathcal{S}_2$ is determined. The surface $\mathcal{S}_2$ could be, for instance, a 1-parameter subcongruence.

Although the amount of Cauchy data is much larger than one worldline and the vorticity of the congruence on that line, as it happens in the Newtonian case, we have a way of getting a congruence out of a part of it on the basis of a ``rigidity'' condition.

In the particular case that one of the worldlines in the congruence is a
geodesic, and the (2+1)-spacetime is flat, reference \cite{LLOSA97} goes a
little further: given the congruence's vorticity on the geodesic and
assuming that strain vanishes on that worldline\footnote{This condition has been introduced in reference \cite{BEL94} as an enhancement of Einstein's equivalence principle and has been named {\it geodesic equivalence principle} \cite{LLOSA97}.}, the conformal rigidity condition (\ref{E.5a}) then determines a unique 2-parameter congruence. The latter would be useful to model a disk whose center is at rest (or in uniform
motion), that spins at an arbitrary angular speed, and that remains {\it as rigid as possible.}

Another remarkable result in \cite{LLOSA97} is that it exists a flat, rigid,
spatial metric, $\overline{g}_{\alpha \beta }$, which is conformal to the
radar metric, $\widehat{g}_{\alpha \beta }$.

The fact that the class of {\it conformally rigid} congruences in a (2+1)-spacetime is ``wide enough'' reminds the well know Gauss theorem \cite{EISENHART1}:

\begin{quote}
\noindent Any Riemannian 2-dimensional space can be conformally mapped into a flat space
\end{quote}

This suggests us a way to extend to (3+1)-spacetimes the results derived for (2+1)-spacetimes, namely, to inspire the formulation of meta-rigidity conditions\footnote{The word meta-rigidity was coined in \cite{BEL95a} to generically refer
to any relativistic extension of the notion of rigidity} in some extension of Gauss theorem to Riemannian 3-manifolds. One instance of this is \cite{BEL98}, where Walberer's theorem \cite{WALB33} is taken as the starting point. In the present paper we shall consider the following

\begin{theorem}  \label{T1}
{\bf (Cartan)} {\rm \cite{CARTAN}} Let (M, g) be a Riemannian 3-manifold. There exist local charts of mutually orthogonal coordinates. Moreover, this can be done in an infinite number of ways.
\end{theorem}

This means that six functions, namely, three coordinates $y^i$ and three factors $f_i$, $i=1,2,3$ can be locally found such that the metric coefficients in this local coordinates are 
\begin{equation}  \label{E.5c}
g_{ij}(y)= f^2_i(y)\delta_{ij}  \,, 
\end{equation}
that is, the Riemannian metric locally admits an orthogonal base which is holonomic.

This result suggests us to advance the following definition

\begin{definition}  \label{d1}
A congruence is said to be  radar-holonomic  iff the associated radar metric $\widehat{g}_{\alpha \beta }$ admits an orthogonal base which is holonomic or, equivalently, if it exists a system of coordinates diagonalizing it.
\end{definition}

That is, six functions exist: $y^i,$ $f_i,$ $i=1,2,3$, such that : 
\begin{equation}  \label{E.5d}
\widehat{g}_{\mu \nu } dx^\mu dx^\nu =\sum_{i=1}^3 f^2_i (dy^i)^2
\end{equation}
the summation convention is understood throughout the paper unless the contrary is explicitly indicated (if either one of the repeated indices is in brackets, or both are superindices (resp. subindices), then the convention is suspended in that formula). Greek indices run from 1 to 4 and lattin indices from 1 to 3.

Section 2 is devoted to develop some geometrical properties of radar-holonomic  congruences, and in section 3 the existence of these
congruence is discussed and posed as a Cauchy problem for a partial differential system. As far as the Cauchy-Kowalewski theorem is invoked in section 3, the analyticity of both the Cauchy hypersurface and the spacetime
metric, will be assumed throughout the paper. The method is somewhat similar to that used in proving
the existence of orthogonal triples of coordinates in a Riemannian 3-manifold \footnote{It has been specially inspiring the reading of reference \cite{DETURK84}, although we have not been able to solve the present existence problem at
the $\mathcal{C}^{\infty}$ level}. We must insist in that the results here derived are valid only locally. No global aspect of spacetimes has been considered. In section \ref{S5} we derive a radar-holonomic 3-parameter congruence which contains a given origin worldline with a prescribed vorticity on it, this is intended to be the mathematical description of a meta-rigid motion.

\section{ Radar-holonomic  congruences \label{S1}}
Let ${\cal C}$ be a  radar-holonomic  3-parameter congruence and let $u(x)$
be the unit tangent vector. According to Definition 1, the radar metric, $\widehat g$, can be written as in equation (\ref{E.5d}). Consider the differential 1-forms $\omega ^i=f_{(i)}dy^i \,,$   $i =1,2,3$.
We thus have: 
\begin{equation}  \label{E6b}
\widehat{g}={\delta}_{ij}\omega ^i{\otimes \omega }^j
\end{equation}
Moreover, since $\widehat{g}$ is orthogonal to $\mathbf{u}$, we have that 
\begin{equation}  \label{E6}
{i_\mathbf{u}\omega ^l=0}         \qquad l =1,2,3
\end{equation}
As a consequence, the functions $y^i$ are constant along any worldline in
the congruence: $\mathcal{L}_\mathbf{u}(y^i)=0 $ and they serve as spatial coordinates adapted to the congruence.

Let us now introduce the differential 1-form $\omega ^4 := -g(\mathbf{u},\_)=u_\alpha (x)dx^\alpha \,$.
From (\ref{E.3}) and (\ref{E6b}) it follows that 
\begin{equation}  \label{E7}
g=\widehat{g}-\omega ^4\otimes \omega ^4 = \eta_{\alpha \beta }\omega^\alpha \omega ^\beta
\end{equation}
where $ \eta_{\alpha \beta }:=(+ + + -)$. 

By definition, the 1-forms $\omega^i$ must be integrable or, equivalently, they must satisfy: 
\begin{equation}  \label{E8}
d\omega^i\wedge \omega^i=0                 \qquad i =1,2,3
\end{equation}

As a result we have thus proved the following
\begin{proposition} \label{p1}
Let (${\cal V}_4$, g) be a spacetime and ${\cal C}$ a  radar-holonomic 3-parameter congruence with unit velocity vector $\mathbf{u}$. Then there exist three integrable 1-forms $\omega^i$, i = 1, 2, 3 such that completed with $\omega^4\equiv -g(\mathbf{u},\_)$, yield a $g$-orthonormal frame.
\end{proposition}

The converse can be easily proved too:
\begin{proposition}  \label{p2}
Let \{$\omega^\alpha $\}$_{\alpha =1..4}$ be a $g$-orthonormal frame such
that $\omega^i$, $i=1,2,3$ are spacelike and integrable and let $\{\mathbf{e}_\alpha\}_{\alpha=1\ldots 4}$ be the dual base, then the flow of $u= \mathbf{e}_4$ is a 3-parameter  radar-holonomic  congruence.
\end{proposition}

\subsection{Geometric properties}
We now list some geometric properties  concerning the strain rate tensor of a radar-holonomic congruence. 
\begin{proposition}  \label{P1}
Let $\omega^l\in \Lambda^1({\cal V}_4)$, $l=1,2,3$, be the orthonormal set
fulfilling conditions (\ref{E6}), (\ref{E7}) and (\ref{E8}) above. Then 
\begin{equation}  \label{E9}
\mathcal{L}_\mathbf{u}\omega^l\wedge \omega^l=0
\end{equation}
\end{proposition}

\smallskip\noindent {\bf Proof:} Using (\ref{E6}) and (\ref{E8}) we can
write: 
\begin{eqnarray*}
\mathcal{L}_\mathbf{u}\omega^l\wedge \omega^l & =&
[i_\mathbf{u}d\omega^l+d(i_\mathbf{u}\omega^l)]\wedge \omega^l=i_\mathbf{u}d\omega^l\wedge \omega^l 
\nonumber \\
&=& i_\mathbf{u}[d\omega^l\wedge \omega^l]-d\omega^l\wedge i_\mathbf{u}\omega^l=0       \nonumber 
\end{eqnarray*}
\hfill$\Box$

\begin{proposition}  \label{P2}
The strain rate tensor $\Sigma \equiv \mathcal{L}_\mathbf{u}\widehat{g}$ has $\omega^i, 
$ $i=1,2,3$ as principal directions. Furthermore, the same holds for any of
its Lie derivatives along the congruence: $\Sigma^{(n)}\equiv \mathcal{L}_\mathbf{u}^n\Sigma =\mathcal{L}_\mathbf{u}^{n+1}\widehat{g}$
\end{proposition}

\smallskip\noindent {\bf Proof:} As a consequence of proposition 1, there
exist three functions $\phi_l,$ $l=1,2,3$ such that $\mathcal{L}_\mathbf{u}\omega^l=\phi_{(l)}\omega^l$. Hence 
\[
\Sigma \equiv \mathcal{L}_\mathbf{u}\widehat{g}=2\phi_i\delta_{ij}\omega^i\otimes
\omega^j \,. 
\]
The second statement, concerning $\Sigma^{(n)}\,$, is easily shown by
induction. \hfill$\Box$

A sort of converse result is the following

\begin{proposition}  \label{P3}
If $\mathcal{L}_\mathbf{u}\Sigma $ and $\Sigma $ diagonalize in the same $g$-orthonormal base, then there exists an orthonormal set $\omega^l\in \Lambda^1({\cal V}_4)$, $l=1\ldots3$, such that 
\begin{equation}  \label{E9b}
i_\mathbf{u}\omega^l = 0 \qquad {\rm and} \qquad 
\mathcal{L}_\mathbf{u}\omega^l\wedge \omega^l=0
\end{equation}
\end{proposition}

\smallskip\noindent
{\bf Proof:}
According to the hypothesis there exist three 1-forms $\rho^i,\;i=1,2,3$ such that 
\begin{eqnarray}
\label{E10a}
\widehat{g}=\delta_{ij}\rho^i\otimes \rho^j \,, \qquad &
\Sigma =2\phi_i\delta_{ij}\rho^i\otimes \rho^j  \\
\label{E10b}
  &\mathcal{L}_\mathbf{u}\Sigma =2\psi_i\delta_{ij}\rho^i\otimes \rho^j 
\end{eqnarray}
These $\rho^i$'s are orthogonal to $\mathbf{u}$, and the same holds for $\mathcal{L}_\mathbf{u}\rho^l$, hence: 
\begin{equation}  \label{E10c}
\mathcal{L}_\mathbf{u}\rho^j=A_{\;k}^j\rho^k
\end{equation}

This can be used to calculate the Lie derivatives of (\ref{E10a}) and, comparing them with (\ref{E10a}) and (\ref{E10b}), we obtain:
\begin{equation}  \label{E11a}
(A_{\;i}^j+A_{\;j}^i) = 2\phi_{(i)}\delta_{ij} \,,  \qquad
\dot{\phi }_{(i)}\delta_{ij}+\phi_{(i)}A_{\;j}^i+
\phi_{(j)}A_{\;i}^j) = \psi_{(i)}\delta_{ij} 
\end{equation}
From which it easily follows that: 
\begin{eqnarray}   \label{E12}
 A_{\;i}^i=\phi_i & , \qquad  &\qquad A_{\;i}^j=-A_{\;j}^i,\quad i\neq j \\
\mbox{for } i=j: &\qquad & \dot{\phi_i}+2\phi_i^2=\psi_i  \nonumber \\   \label{E13}
\mbox{for } i\neq j: &\qquad & (\phi_{(i)}-\phi_{(j)})A_{\;i}^j=0 
\end{eqnarray}

Now three cases must be considered according to the degeneracy of the
eigenvalues of $\Sigma $.
\begin{list}
{\alph{llista})}{\usecounter{llista}}
\item In the case $\phi_1\neq \phi_2\neq \phi_3$ the set $\{\rho^i\}$ is
      unambiguously defined. Eq.(\ref{E13}) implies $A_{\;i}^j=0,i\neq j.$
      Then taking  
     (\ref{E12}) and (\ref{E10c}) into account we obtain $\mathcal{L}_\mathbf{u}\rho^i=
     \phi_{(i)}\rho^i$ and equation (\ref{E9b}) follows for $\omega^i=\rho^i$.
\item {In the case $\phi_1=\phi_2\neq \phi_3$ from (\ref{E12}) and
(\ref{E13}), we obtain:
\begin{equation}
\label{E14a}
\left.
\begin{array}{ll}
A_{\;i}^i=\phi_i \,,\qquad & i=1,2,3 \,; \\
A_{\;2}^1=-A_{\;1}^2  & 
A_{\;a}^3=-A_{\;3}^a=0 \,,\qquad a=1,2 
\end{array}
\right\}
\end{equation}
which introduced in (\ref{E10c}) yield: 
\begin{equation}
\label{E14b}
\mathcal{L}_\mathbf{u}\rho^3=\phi_{(3)}\rho^3\qquad ;\qquad \mathcal{L}
_\mathbf{u}\rho^a=A_{\;b}^a\rho^b\qquad a,b=1,2 
\end{equation}

In the present case, however, the set $\omega^3=\rho^3 \quad \mbox{and }\quad \omega^a=R_{\;b}^a\rho^b\qquad a,b=1,2 
$ with $(R_{\;b}^a)\in O(2)$ is also a set of eigenvectors for 
$\Sigma $. Now, an orthogonal matrix $(R_{\;b}^a)$ can be found such that
$\mathcal{L}_\mathbf{u}\omega^a=\phi \omega^a$, with $\phi_1=\phi_2=\phi $. 
Indeed, since $\mathcal{L}_\mathbf{u}\omega^a=[\mathcal{L}_\mathbf{u}R_{\;b}^a(R^{-1})_{\;c}^b+ (RAR^{-1})_{\;c}^a]\omega^c $
it is enough to require 
$$\mathcal{L}_\mathbf{u}R_{\;b}^a=R_{\;c}^a(-A_{\;b}^c+\phi \delta_{\;b}^c)$$ 
which has many solutions $(R_{\;b}^a)\in O(2)$ because, by equation (\ref{E14a}),
$-A_{\;b}^c+\phi\,\delta_b{\;}^c\,$ is skewsymmetric.}
\item The completely degenerate case $\phi_1=\phi_2=\phi_3$ can be handled
in much the same way as case (b).\hfill$\Box$
\end{list}

\begin{theorem}   \label{T4}
$\mathcal{L}_\mathbf{u}\Sigma $ and $\Sigma $ diagonalize in a common $g$-orthonormal
base if, and only if, three functions $A$, $B$ and $C$ exist such that 
\begin{equation}
\label{E15}\mathcal{L}_\mathbf{u}\Sigma =A\widehat{g}+B\Sigma +C\Sigma^2 
\end{equation}
where $\Sigma_{\alpha \beta }^2\equiv\Sigma_{\alpha \mu
}^{}\Sigma_{\;\beta}^\mu $.  Moreover if two among the eigenvalues of
$\Sigma $ are equal, then $C=0$ can be taken, and in the completely
degenerate case, $B=C=0$ can be taken.
\end{theorem}

\smallskip\noindent
{\bf Proof:} \\
{\boldmath $(\Rightarrow)$:}\hspace*{.5em} 
Since $\mathbf{u}$ is orthogonal to both $\Sigma $ and $\mathcal{L}_\mathbf{u}\Sigma $, 
we shall have that $\omega^4\equiv -g(u,\_)$ is in the common orthogonal 
base $\{\omega^\alpha \}$. Thus, expressions similar
to (\ref{E10a}) and (\ref{E10b}) hold. Hence to prove (\ref{E15}) amounts to
solve the linear system: 
\begin{equation}
\label{E16}A+2\phi_iB+4\phi_i^2C=2\psi_i 
\end{equation}
for the unknowns $A$, $B$ and $C$. The determinant is:
$$
\Delta =8(\phi_2-\phi_1)(\phi_2-\phi_3)(\phi_3-\phi_1) \,.
$$
In the non-degenerate case $\Delta \neq 0$ and (\ref{E16}) has a unique
solution.

If $\phi_1=\phi_2\neq \phi_3$, only the equations for $l=2$ and $3$ in 
(\ref{E16}) are independent, there are infinitely many solutions and $C$ can
be arbitrarily chosen. In particular, $C=0$.

Finally, in the completely degenerate case, (\ref{E16}) has rank 1, hence
it admits infinitely many solutions, and $B$ and $C$ are arbitrary.

\smallskip\noindent
{\boldmath $(\Leftarrow)$:}\hspace*{.5em} Assume that (\ref{E15}) holds. Since $\Sigma $ diagonalize in a
$g$-orthonormal base, we substitute (\ref{E10a}) in (\ref{E15}) and it
follows immediately that $\mathcal{L}_\mathbf{u}\Sigma $ diagonalize in the 
same $g$-orthonormal base.

\begin{theorem}  \label{T5}
If $\Sigma $, $\mathcal{L}_\mathbf{u}\Sigma $ and $\mathcal{L}_\mathbf{u}^2\Sigma $ diagonalize in
a common $g$-orthonormal base, then $\widehat{g}$, $\Sigma $, 
$\mathcal{L}_\mathbf{u}\Sigma $ and $\mathcal{L}_\mathbf{u}^2\Sigma $ are linearly dependent.
(Hence,  the congruence is non-generic \cite{BEL91}.)  
\end{theorem}

\smallskip\noindent
{\bf Proof:}
By theorem \ref{T4}, there exist $A$, $B$ and $C$ such that (\ref{E15})
holds. taking the Lie derivative on both sides, using that 
$\Sigma =\mathcal{L}_\mathbf{u}\widehat{g}$ and equation (\ref{E15}) itself, and taking into
account that the minimal polynomial for $\Sigma_\beta^\alpha \,$ has at
most degree 3, we arrive at: 
\begin{equation}
\label{E17}\mathcal{L}_\mathbf{u}^2\Sigma =A^{\prime }\widehat{g}+B^{\prime }\Sigma
+C^{\prime }\Sigma^2 
\end{equation}
where $A^{\prime }$, $B^{\prime }$ and $C^{\prime }$ are some suitable
functions.
If $C=0$, then (\ref{E15}) already proves the theorem and, if on the contrary $C\neq 0$, we can derive $\Sigma^2$ from (\ref{E15}) and substitute it into (\ref{E17}), so arriving at:
$$
\mathcal{L}_\mathbf{u}^2\Sigma =\left(A^{\prime }-\frac AC\right) \,\widehat{g} +\left(B^{\prime }-\frac BC\right)\,\Sigma + \frac{C^{\prime }}C \,\mathcal{L}_\mathbf{u}\Sigma  \hspace*{4em}\Box $$

\section{Existence of  radar-holonomic  congruences \label{S3}}
According to the propositions 1 and 2 in section \ref{S1}, proving the existence of radar-holonomic  congruences is equivalent to prove the existence of a
$g$-orthonormal base $\{\omega^\alpha \}$ such that:

\begin{list}
{(\roman{llista})}{\usecounter{llista}}
\item \quad $\omega^i$ is spacelike, and
\item \quad $d\omega^i\wedge \omega^i=0$ \hfill (\ref{E8})
\end{list}
The dual tetrad will be denoted $\{\mathbf{e}_\alpha \}$ and the commutation
relations: 
\begin{equation}
\label{e18}
[\mathbf{e}_\alpha ,\mathbf{e}_\beta ]=\,C_{\alpha \beta }^\mu \mathbf{e}_\mu \qquad 
d\omega^\alpha =-\frac 12\,C_{\mu \nu }^\alpha \,\omega^\mu \wedge \omega^\nu 
\end{equation}
Using this, equation (\ref{E8}) can be written as: 
\begin{equation}
\label{e19}
-\frac 12\,C_{\mu \nu }^i\omega^\mu \wedge \omega^\nu \wedge
\omega^i=0 \qquad i=1,2,3
\end{equation}
which in turn is equivalent to: 
\begin{equation}
\label{e19b}
C_{jk}^i=0\quad {\rm for} \quad i\neq j\neq k \qquad {\rm and} \qquad  
C_{4j}^i=0\quad {\rm for} \quad i\neq j 
\end{equation}

Taking into account the relationship between $C_{\alpha \beta }^\gamma $ and the Levi-Civita connection coefficients in an orthonormal frame \cite{CHOQUET}, equations (\ref{e19b}) are equivalent to:  
\begin{equation}   \label{e20a}
 \gamma_{jk}^i  =0 \quad {\rm for} \quad i\neq j\neq k \qquad {\rm and} \qquad 
C_{4j}^i  =\gamma_{4j}^i-\gamma_{j4}^i=0 \quad {\rm for} \quad   i\neq j 
\end{equation}

As the connection symbols in an orthonormal frame fulfill that $\gamma_{\alpha k}^i=-\gamma_{\alpha i}^k$ and  $\gamma_{\alpha k}^4=\gamma_{\alpha 4}^k$, at most three among the equations (\ref{e20a}) for $i\neq j \neq k$ are independent.

Since $\gamma_{\alpha \beta }^\gamma =(\omega^\gamma ,\nabla_\alpha \mathbf{e}_\beta )=\eta^{\gamma \nu }g(\mathbf{e}_\nu ,\nabla_\alpha \mathbf{e}_\beta ) $, equations (\ref{e20a}) yield a first-order partial differential system of nine equations where the unknown is the orthonormal frame $\{\mathbf{e}_\alpha \}$.

\subsection{The Cauchy problem}
Given a hypersurface ${\cal S}_0$, let $\mathbf{n}$ be a unit vector field such that, when restricted to ${\cal S}_0$, is orthogonal to it and that the 1-form $\nu =g(\mathbf{n},\_)$ is integrable, i.e. $\mathbf{n}$ is hypersurface orthogonal. 

Relatively to $\mathbf{n}$, each vector $\mathbf{e}_\alpha $ of the sought frame can be decomposed in an orthogonal part $\mathbf{e}_\alpha^\top$ (which is tangent to the hypersurfaces ${\cal S}_\lambda$) and a parallel part, namely: 
\begin{equation}
\label{e23}
\mathbf{e}_\alpha =\mathbf{e}_\alpha^\top + n_\alpha \mathbf{n} \,, \qquad {\rm where} \qquad n_\alpha \equiv g(\mathbf{e}_\alpha ,\mathbf{n})
\end{equation}
As $\mathbf{n}$ is hypersurface orthogonal, they exist local charts $(y^1, \ldots y^4)$ such that 
$$\mathbf{n}= \frac{\partial}{\partial y^1} \qquad {\rm and} \qquad  \mathbf{e}_\alpha^\top y^1 = 0 $$

Substituting (\ref{e23}) into (\ref{e20a}) and using that $\gamma_{\alpha \beta }^\gamma = \eta^{\gamma \nu} g(\mathbf{e}_\nu ,\nabla_\alpha \mathbf{e}_\beta ) $, we arrive at: 
\begin{eqnarray}
\label{e24a}
\gamma_{jk}^i \equiv n_j W_{ik}+ g(\mathbf{e}_i, \nabla^\top_j \mathbf{e}_k)=0 &\qquad & i\neq j\neq k  \\
\label{e24b}
C_{4k}^i \equiv n_4W_{ik}-n_kW_{i4}+H_{ik}=0 &\qquad & i\neq k 
\end{eqnarray}
where $\nabla^\top_j \mathbf{e}_k$ is the covariant derivative along $\mathbf{e}_j^\top$,
\begin{equation}
\label{e25a}
W_{\alpha \beta }:= g(\mathbf{e}_\alpha ,\nabla_\mathbf{n} \mathbf{e}_\beta ) 
\end{equation}
is skewsymmetric, and 
\begin{equation}    \label{e25}
H_{ik}:=  g(\mathbf{e}_i ,\nabla^\top_4 \mathbf{e}_k)- g(\mathbf{e}_i ,\nabla^\top_k \mathbf{e}_4)
\end{equation}
only depends on the unknowns and their derivatives along directions which are orthogonal to $\mathbf{n} $. 

The principal part of the partial differential system (\ref{e24a}-\ref{e24b}), i. e. the terms containing derivatives along $\mathbf{n}$, is all included in the $W_{\alpha\beta}$ terms. Besides, there are three combinations of these equations 
\begin{equation}  \label{e27}
S^i := 2 C^i_{4[k} n_{j]} - 2  n_4 \gamma^i_{[jk]} =0
\end{equation}
which do not contain normal derivatives. This is obvious if we realise that 
$$ S^i =\frac12\, C^i_{\alpha\beta} n_\lambda \epsilon^{\alpha\beta\lambda i} = \langle \omega^i, n_\lambda \epsilon^{\alpha\beta\lambda i} \nabla_\alpha \mathbf{e}_\beta \rangle =g(\mathbf{e}_i, n_\lambda \epsilon^{\alpha\beta\lambda i} \nabla_\alpha^\top \mathbf{e}_\beta \rangle $$

If, and only if, $n_j\neq 0$, the nine equations (\ref{e24a}) and (\ref{e24b}) can be solved for the six independent components of $W_{\alpha \beta }$. A little algebra yields: 
\begin{equation}   \label{e26a}
\left. \begin{array}{ll}
\displaystyle{W_{ik} = -\frac 1{n_j} g(\mathbf{e}_i,\nabla^\top_j\mathbf{e}_k)}  & \quad i\neq j\neq k \\ 
\displaystyle{W_{i4} = \frac 1{n_k}H_{ik}-\frac{n_4}{n_k n_j} g(\mathbf{e}_i,\nabla^\top_j\mathbf{e}_k)}  & \quad i\neq j\neq k 
\end{array}  \right\}
\end{equation}
Notice that for each value of $i=1,2,3$ there are two ways of choosing $j\neq k$ in the second equation (\ref{e26a}). This comes from the fact that the system (\ref{e24a})-(\ref{e24b}) is overdetermined. It can be easily seen that compatibility implies the  subsidiary conditions (\ref{e27}), which only depend on ``tangential derivatives'' of the unknowns and are to be taken as constraints on the Cauchy data on $\mathcal{S}_0$.

As $\displaystyle{\nabla_\mathbf{n} \mathbf{e}_k = \sum_{i=1}^3 W_{ik} \mathbf{e}_i - W_{4k} \mathbf{e}_4}$, the above equation yields 
\begin{equation}  \label{e28a}
\nabla_\mathbf{n} \mathbf{e}_k = - \sum_{i\neq j\neq k=1}^3 \frac1{n_j} g(\mathbf{e}_i,\nabla^\top_j\mathbf{e}_k)
\mathbf{e}_i + \left(  \frac 1{n_l}H_{kl}-\frac{n_4}{n_l n_j} g(\mathbf{e}_k,\nabla^\top_j\mathbf{e}_l) 
\right) \mathbf{e}_4
\end{equation}
where $(lkj)$ is a cyclic permutation of $(123)$.
Furthermore, due to the orthonormality, the $\mathbf{n}$-covariant derivative of $\mathbf{e}_4$ is determined too. Hence, if the orthonormal frame $\{\mathbf{e}_\alpha\}_{\alpha=1\ldots 4}$ is given on $\mathcal{S}_0$, it can be determined on a neighbourhood of $\mathcal{S}_0$.  

\begin{lemma} \label{L1}
Given an orthonormal tetrad $\{\overline{\mathbf{e}}_\alpha \}$ on ${\cal S}_0$,
such that $n_i:=g(\mathbf{n},\overline{\mathbf{e}}_i)\neq 0$, it exists a neighbourhood ${\cal U}$ of 
${\cal S}_0$ and an orthonormal tetrad $\{\mathbf{e}_\alpha \}$ such that $\mathbf{e}_\alpha =
\overline{\mathbf{e}}_\alpha $ on ${\cal S}_0$ and is a solution of (\ref{e28a}).
\end{lemma}

\smallskip\noindent
{\bf Proof:}
Let $\{\widetilde{\mathbf{e}}_\alpha \}$ be a given orthonormal frame. In terms of it, the unknown frame $\{\mathbf{e}_\alpha \}$ can be written as $ \mathbf{e}_\alpha =L_{\ \alpha }^\mu \widetilde{\mathbf{e}}_\mu $,
where $(L_{\ \alpha }^\beta)$ is a Lorentz matrix valued function.

Substituting the latter into equation (\ref{e28a}) we have:
\begin{equation}  \label{e28b}
 \mathbf{n}L_{\ \nu }^\alpha + \eta^{\rho \nu } L_{\ \alpha }^\mu 
g(\widetilde{\mathbf{e}}_\rho ,\nabla_\mathbf{n} \widetilde{\mathbf{e}}_\mu ) = {\rm n.p.t.}
\end{equation}
where ``n.p.t.'' means {\em non-principal terms}, that is terms containing derivatives along $\mathbf{e}_\alpha^\top$.
In the coordinates $(y^1 \ldots y^4)$ adapted to $\mathbf{n}$ mentioned at the beginning of the section, these terms depend at most on $\partial L^\alpha_{\ \nu}/\partial y^\rho$, $\rho=2,3,4$, whereas $\mathbf{n}L_{\ \nu }^\alpha = \partial L^\alpha_{\ \nu}/\partial y^1$. Hence, (\ref{e28b}) is already in normal form and the Cauchy-Kowalevski theorem \cite{JOHN71} implies that, given $L^\alpha_{\ \nu} (0,y^2,y^3,y^4)$, i. e. the values of $L^\alpha_{\ \nu}$ on $\mathcal{S}_0$, a solution of (\ref{e28b}) exists in a neigbourhood of the Cauchy hypersurface.  \hfill $\Box$

\begin{lemma} \label{L2}
Let $\{\mathbf{e}_\alpha \}$ be an orthonormal tetrad, which is a solution of (\ref{e28a}) and fulfills the subsidiary conditions (\ref{e27}) on ${\cal S}_0$. Then (\ref{e27}) holds in the neighbourhood ${\cal U}$ of ${\cal S}_0$ where $\{\mathbf{e}_\alpha \}$ is defined.
\end{lemma}

\smallskip\noindent
{\bf Proof:}
Since $\{\mathbf{e}_\alpha \}$ is a solution of (\ref{e28a}), the $W_{i\alpha}$ are given by (\ref{e26a}) and, using (\ref{e27}), we easily obtain that:
\begin{equation}
\label{e29}C_{jk}^i=0\, ,\qquad C_{4k}^i  =0\qquad {\rm and} \qquad C_{4j}^i n_k =  - S^i 
\end{equation}
where $(ikj)$ is a cyclic permutation of $(123)$ and $S_i$ is defined in (\ref{e27}).

The commutation coefficients $C_{\alpha \beta }^\mu $ satisfy the Jacobi identity: 
$\epsilon^{\nu \rho \alpha\beta }(\mathbf{e}_\rho C_{\alpha\beta}^\mu - C_{\sigma \rho }^\mu C_{\alpha\beta}^\sigma)=0 $.
Taking $\mu =\nu =i$ and using (\ref{e29}) and the decomposition (\ref{e23}), we arrive at:
$$ \mathbf{e}_k C^i_{4j} + C_{4j}^i  C_{k\alpha}^\alpha  = 0$$
where $(ikj)$ is a cyclic permutation of $(123)$. 
This can be considered as a linear, first order, partial differential system on $C^i_{4j}$. If $n_k\neq 0$, the hypersurface ${\cal S}_0\,$ is non-characteristic and the above equation has a unique analytic solution for $S^i=0$ on ${\cal S}_0\,$; this ensures that $C^i_{4j}=0$, and hence $S^i=0$ in a neighbourhood of ${\cal S}_0\,$.

Summarizing, we have so far proved that 

\begin{theorem} \label{T6}
Given an orthonormal tetrad $\{\overline{\mathbf{e}}_\alpha \}$ on ${\cal S}_0$, such that: {\em (i)}  $\; g(\mathbf{n},\overline{\mathbf{e}}_i)\neq 0\quad $ and {\em (ii)} $\; S_i=0\quad $ on ${\cal S}_0$. Then it exists an orthonormal frame $\{\mathbf{e}_\alpha \}$ in a neighbourhood ${\cal U}$ of ${\cal S}_0$, such that:
\begin{list}
{(\roman{llista})}{\usecounter{llista}}
\item it is a solution of (\ref{e28a}) and (\ref{e27}), and
\item $\quad \mathbf{e}_\alpha =\overline{\mathbf{e}}_\alpha \quad $ on $\,{\cal S}_0$
\end{list}
Thus, the congruence generated by $\mathbf{u}=\mathbf{e}_4$ is  radar-holonomic .
\end{theorem}

The proof is straightforward from lemmas \ref{L1} and \ref{L2}. \hfill $\Box$

\section{Getting a  radar-holonomic  3-congruence out of a given 2-congruence \label{S4}}
We now choose the hypersurface ${\cal S}_0\,$ so that it is spanned by a given 2-congruence of worldlines. $\overline{\mathbf{u}}$ and $\overline{\mathbf{n}}$ will denote the unit velocity vector for the 2-congruence and the unit vector orthogonal to ${\cal S}_0$, respectively. (Hereafter, a bar over a symbol indicates that we are only considering the values of that object on ${\cal S}_0$.)

\subsection{Are the subsidiary conditions consistent?}
We are interested in finding an orthonormal tetrad $\{\overline{\mathbf{e}}_\alpha \}$ on ${\cal S}_0$, such that: 
$\overline{\mathbf{e}}_4=\overline{\mathbf{u}}$, \,$\overline{n}_i\equiv g(\overline{\mathbf{e}}_i, \overline{\mathbf{n}})\neq 0\,$, and that $S_i=0$ on ${\cal S}_0$.

We shall first see that the giving of the orthonormal triad $\{\overline{\mathbf{e}}_i \}_{i=1\ldots 3}$ is equivalent to the giving of the three directions $\{\widehat{\mathbf{a}}_i \}$ of their tangential projections on $\mathcal{S}_0$. We shall then see that $S_i=0$ amounts to three conditions on these directions and also that there exist triples of such directions $\{\widehat{\mathbf{a}}_i \}$ fulfilling these conditions.

For each $\overline{\mathbf{e}}_i$, we consider the decomposition (\ref{e23}) 
and take the unit vector 
\begin{equation}
\label{e34}\widehat{\mathbf{a}}_i\equiv \frac{\overline{\mathbf{e}}_i^\top}{\left\| 
\overline{\mathbf{e}}_i^\top\right\| } 
\end{equation}

Furthermore, we can consider the combinations: 
\begin{equation}
\label{e35}
\overline{\mathbf{b}}_i=\epsilon_i^{\ jk}\overline{n}_j\,\overline{\mathbf{e}}_k \qquad
\qquad i=1,2,3 
\end{equation}
and define the unit vector $\widehat{\mathbf{b}}_i\equiv \overline{\mathbf{b}}_i/\|\overline{\mathbf{b}}_i\|$.
It is straightforward to see that $\{\overline{\mathbf{u}},\overline{\mathbf{n}},\widehat{\mathbf{b}}_i,
\widehat{\mathbf{a}}_i\}$ is an orthonormal tetrad at any point in ${\cal S}_0$.

We also have that 
$$g(\widehat{\mathbf{a}}_i,\widehat{\mathbf{a}}_j)= \frac 1{\left\| 
\overline{\mathbf{e}}_i^\top\right\| \left\| \overline{\mathbf{e}}_j^\top\right\|}\, g(\overline{\mathbf{e}}_i^\top, \overline{\mathbf{e}}_j^\top)= \frac 1{\left\|\overline{\mathbf{e}}_i^\top \right\| \left\| \overline{\mathbf{e}}_j^\top\right\| }\,(\delta_{ij}-\overline{n}_i \overline{n}_j)$$ 
Now, since $\overline{n}_i\neq 0$, $\forall i$, then $\left| \overline{n}_j\right| <1$, $\forall i$,and
it follows that:
\begin{equation}
\label{e36}
\forall i\neq j\left| g(\widehat{\mathbf{a}}_i,\widehat{\mathbf{a}}_j)\right| <1\qquad 
{\rm and} \qquad 
g(\widehat{\mathbf{a}}_i,\widehat{\mathbf{a}}_j)\neq 0 
\end{equation}

Thus, for any triad $\{\overline{\mathbf{e}}_1,\overline{\mathbf{e}}_2,\overline{\mathbf{e}}_3\}$ in 
$T_x{\cal V}_4$, $x \in{\cal S}_0$, such that: 
$g(\overline{\mathbf{e}}_i,\overline{\mathbf{u}})=0$, and 
$g(\overline{\mathbf{e}}_i,\overline{\mathbf{n}})=\overline{n}_i\neq 0$, we can obtain a triad 
$\widehat{\mathbf{a}}_i\in $ $T_x{\cal S}_0$, $i=1,2,3$, such that
$g(\widehat{\mathbf{a}}_i,\overline{\mathbf{u}}) =0$ and that (\ref{e36}) holds.

The converse result is the following

\begin{proposition}   \label{p3}
Given a triad $\widehat{\mathbf{a}}_i\in $ $T_x{\cal S}_0$, $i=1,2,3$, 
such that $g(\widehat{\mathbf{a}}_i,\overline{\mathbf{u}})=0$ and fulfills (\ref{e36}), 
a triad $\{\overline{\mathbf{e}}_i\}$ can be obtained such that
$g(\overline{\mathbf{e}}_i,\overline{\mathbf{n}})\neq 0\,$ and that
$\{\overline{\mathbf{e}}_1,\overline{\mathbf{e}}_2,\overline{\mathbf{e}}_3,\overline{\mathbf{e}}_4=
\overline{\mathbf{u}}\}$ is an orthonormal frame.
\end{proposition}

\smallskip\noindent
{\bf Proof:}
We must find $A_i$ and $B_i$ such that $\overline{\mathbf{e}}_i=A_i\widehat{\mathbf{a}}_i+B_i\overline{\mathbf{n}}$ and,
since $\widehat{\mathbf{a}}_i$ and $\overline{\mathbf{n}}$ are $\mathbf{u}$-orthogonal, so will be 
$\overline{\mathbf{e}}_i$.

Now, the condition $g(\overline{\mathbf{e}}_i,\overline{\mathbf{e}}_j)=\delta_{ij}$ implies that 
$$ A_i^2+B_i^2=1 \qquad {\rm and} \qquad 
A_iA_jg(\widehat{\mathbf{a}}_i,\widehat{\mathbf{a}}_j)+B_iB_j=0\qquad i\neq j  \,, $$
from which we easily obtain 
\begin{equation}
\label{e37}B_i=A_i\sqrt{-\frac{g(\widehat{\mathbf{a}}_i,\widehat{\mathbf{a}}_j) \cdot 
g(\widehat{\mathbf{a}}_i,\widehat{\mathbf{a}}_k)}{g(\widehat{\mathbf{a}}_j,\widehat{\mathbf{a}}_k)}}\qquad i\neq
j\neq k 
\end{equation}
and 
\begin{equation}
\label{e38}A_i=\sqrt{\frac{g(\widehat{\mathbf{a}}_j,\widehat{\mathbf{a}}_k)}{g(\widehat{\mathbf{a}}_j,
\widehat{\mathbf{a}}_k)-g(\widehat{\mathbf{a}}_i,\widehat{\mathbf{a}}_j)\cdot g(\widehat{\mathbf{a}}_i,
\widehat{\mathbf{a}}_k)}}\qquad i\neq j\neq k 
\end{equation}
The right hand side of (\ref{e37}) is well defined because, by the
hypothesis, the inequalities (\ref{e36}) hold.

The denominator in (\ref{e38}) neither vanishes, as a consequence of 
(\ref{e36}) too. Indeed, denoting by 
$\varphi_{ij}\,$ the angle between $\widehat{\mathbf{a}}_i$ and $\widehat{\mathbf{a}}_j$, and
taking into account that $\left| \varphi_{ij}\right| +\left| \varphi
_{jk}\right| +\left| \varphi_{ki}\right| =2\pi$, $\, i\neq j\neq k\,$
this denominator reads:
$$
\cos (\varphi_{jk})-\cos (\varphi_{ij})\cdot \cos (\varphi_{ik}) =
\sin (\varphi_{ij})\cdot \sin (\varphi_{ik}) \qquad i\neq j\neq k 
$$
which does not vanish because $\left| \cos (\varphi_{ij})\right|= \left| g(\widehat{\mathbf{a}}_i, \widehat{\mathbf{a}}_j) \right| <1\,,\quad \forall i\neq j$. \hfill$\Box$

We shall now find $\widehat{\mathbf{a}}_1,\widehat{\mathbf{a}}_2,\widehat{\mathbf{a}}_3\,$ 
such that the triad $\{\overline{\mathbf{e}}_i\}$  so reconstructed fulfills the subsidiary condition (\ref{e27}). 
It is easily seen that $S_i$ can be written as:
$$ S_i\equiv n_k g({\mathbf{e}}_i,[{\mathbf{e}}_4,{\mathbf{e}}_j])- n_j g({\mathbf{e}}_i,[{\mathbf{e}}_4, {\mathbf{e}}_k]) = - g({\mathbf{e}}_i,[{\mathbf{e}}_4,\mathbf{b}_i]) $$
[where $(ikj)$ is a cyclic permutation of $(123)$] on the hypersurface ${\cal S}_0$, that is:
$$ \overline{S}_i= - g(\overline{\mathbf{e}}_i,[\overline{\mathbf{u}},\overline{\mathbf{b}}_i]) $$

Taking now into account that $\overline{\mathbf{u}}\,$ and $\overline{\mathbf{b}}_i\,$ are tangential to ${\cal S}_0$ and, therefore, $[\overline{\mathbf{u}},\overline{\mathbf{b}}_i]$ is also tangential, we can write: 
$$ -\overline{S}_i= \left\| \overline{\mathbf{e}}_i^\top\right\| \cdot \left\| \overline{\mathbf{b}}_i\right\| g(\widehat{\mathbf{a}}_i, [\overline{\mathbf{u}},\widehat{\mathbf{b}}_i]) $$
and therefore the subsidiary conditions $\overline{S}_i=0$ are equivalent to 
\begin{equation}
\label{e39}
g(\widehat{\mathbf{a}}_i,[\overline{\mathbf{u}},\widehat{\mathbf{b}}_i]) =0
\end{equation}

Let $\overline{\mathbf{v}}$ and $\overline{\mathbf{m}}$ be two unit vector fields such that
$\{\overline{\mathbf{u}}, \overline{\mathbf{m}}, \overline{\mathbf{v}}\}$ is an orthonormal base on
the tangent space of ${\cal S}_0$. In terms of this base, we can write:
\begin{equation} \label{e39b}
\widehat{\mathbf{b}}_i=\cos \theta_i\,\overline{\mathbf{m}}+\sin \theta_i\,\overline{\mathbf{v}} \qquad {\rm and} \qquad
\widehat{\mathbf{a}}_i=-\sin \theta_i\,\overline{\mathbf{m}}+\cos \theta_i\,\overline{\mathbf{v}} 
\end{equation}
and the conditions (\ref{e39}) lead to: 
$$\dot\theta_i + \frac14 \sin 2\theta_i\, (\overline\Sigma_{\mathbf{v}\mathbf{v}} - \overline\Sigma_{\mathbf{m}\mathbf{m}}) + \frac12 \cos 2\theta_i\,\overline\Sigma_{\mathbf{m}\mathbf{v}} + \frac12\, \left(g(\overline{\mathbf{v}},[\overline{\mathbf{u}},\overline{\mathbf{m}}]) -
g(\overline{\mathbf{m}},[\overline{\mathbf{u}},\overline{\mathbf{v}}])\right) = 0   $$
which is an ordinary differential equation for each $\theta_i$, $i=1,2,3$, and has a solution for every initial data $\theta_i^0$ given in a submanifold ${\cal M}\subset{\cal S}_0$, such that it is nowhere tangent to
$\overline{\mathbf{u}}$. 

Summarizing, we have thus proved that given a 2-congruence, it spans a hypersurface ${\cal S}_0$ on which Cauchy data $\{\overline{\mathbf{e}}_1, \overline{\mathbf{e}}_2,\overline{\mathbf{e}}_3, \overline{\mathbf{e}}_4=\overline{\mathbf{u}}\}$ can be found fulfilling the subsidiary conditions (\ref{e27}). Moreover, this can be done in an infinite number of ways.

\subsection{The kinematical meaning of the Cauchy data}
In the particular case considered in this section, where the Cauchy hypersurface ${\cal S}_0$ is spanned by a 2-congruence, we shall analyse the kinematical significance of the Cauchy data $\{\overline{e}_i\}_{i=1,2,3}$.
According to Proposition \ref{P3}, the latter is a principal basis for $\overline\Sigma$ (i. e., the values of the strain rate of the radar-holonomic 3-congruence. We shall see that  $\{\overline{e}_i\}_{i=1,2,3}$ determine $\overline\Sigma$.

Let $\jmath: \mathcal{S}_0 \hookrightarrow \mathcal{V}_4$ be the canonical embedding and consider: $\overline\Sigma^0:= \jmath^\ast\overline\Sigma$ and $\widehat g^0:= \jmath^\ast\widehat g$, that is
$$\overline\Sigma^0 (\overline{v},\overline{w}) = \overline\Sigma(\overline{v},\overline{w})\,,   \qquad  \widehat g^0(\overline{v},\overline{w}) =\widehat g(\overline{v},\overline{w}) \,, \qquad\forall \overline{v},\overline{w}\in T\mathcal{S}_0 $$ 
It is easy to prove that $\overline\Sigma^0  = \mathcal{L}_{\overline{\mathbf{u}}} \widehat g^0 $, therefore $\overline\Sigma^0$ is the strain rate of the given 2-congruence and is determined by the Cauchy data on $\mathcal{S}_0$ only.

\begin{proposition}\label{P6}
The Cauchy data $({\cal S}_0, \overline{e}_1, \overline{e}_2, \overline{e}_3, \overline{e}_4=\overline{u})$ determine $\overline\Sigma$, i. e., the values on ${\cal S}_0$ of the strain rate of the radar-holonomic congruence obtained as a solution of the partial differential system (\ref{e24a})--(\ref{e24b}).  
\end{proposition}

\paragraph{Proof:} Let $\{\omega^\alpha\}$ be the dual base of $\{e_\mu\}$. By proposition \ref{P2} we have that  
\begin{equation} \label{RT110}
\overline\Sigma = 2\overline\phi_i \,\overline\omega^i\otimes \overline\omega^i
\end{equation}
and, using (\ref{e23}), we obtain
$$ \overline\Sigma^0(\overline{\mathbf{e}}_i^\top,\overline{\mathbf{e}}_j^\top) = \overline\Sigma(\overline{\mathbf{e}}_i- \overline{n}_i\overline{\mathbf{n}},\overline{\mathbf{e}}_j- \overline{n}_j\overline{\mathbf{n}}) = 2 \phi_i - 2 \overline{n}_i \overline{n}_j\left(\phi_i + \phi_j +\sum_{l=1}^3 \phi_l (\overline{n}_l)^2 \right) $$
whence it follows that 
$$ 2\phi_i = \overline\Sigma^0(\overline{e}_h^\top,\overline{e}_l^\top) \delta^{lh} + \frac 1{\overline{n}_j
\overline{n}_k}\overline\Sigma^0(\overline{e}_j^\top,\overline{e}_k^\top)\,,  \qquad i \neq j \neq k $$
which determines the principal values of $\overline\Sigma$ in terms of the Cauchy data and, through (\ref{RT110}), determines $\overline\Sigma$ too. \hfill $\Box$

And conversely

\begin{proposition}\label{P7}
Let us be given a 2-congruence of worldlines spanning $\mathcal{S}_0$ and  let $\overline{\mathbf{u}}$, $\overline{\mathbf{n}}$ and $\overline\Sigma$ be respectively the unit tangent vector, the unit vector normal to $\mathcal{S}_0$ and a symmetric tensor field on  $\mathcal{S}_0$ with values on $T\mathcal{V}_4$ such that  $\overline\Sigma(\overline{\mathbf{u}},-)=0 $  none of whose principal vectors is tangential to $\mathcal{S}_0$. Then it determines a set of Cauchy data $({\cal S}_0, \{\overline{e}_i\})$ with $\overline{e}_4 = \mathbf{u}$ such that $\mathcal{S}_0$ is a non-characteristic hypersurface for the partial differential system (\ref{e24a})--(\ref{e24b}) and
 $\overline\Sigma$ is the strain rate of the resulting radar-holonomic congruence.  
\end{proposition}

\paragraph{Proof:} $\overline\Sigma$ has three different principal values. Indeed, if $\phi_1=\phi_2$, there would be a principal vector $\overline\mathbf{v} = m_i \overline\mathbf{e}_1 + m_2 \overline\mathbf{e}_2$ such that $\widehat g(\overline\mathbf{v},\overline\mathbf{n})=0$, i. e. tangential to $\mathcal{S}_0$. 

In this case, there is a unique base of principal vectors, $\{\overline\mathbf{e}_1,\overline\mathbf{e}_2, \overline\mathbf{e}_3\}$ which, as they are not tangential to $\mathcal{S}_0$, also fulfill that $\overline n_i = \widehat g(\overline\mathbf{e}_i,\overline\mathbf{n})\neq 0$. Therefore the hypersurface $\mathcal{S}_0$ is non-charcteristic for the partial differential system (\ref{e24a})--(\ref{e24b}). \hfill $\Box$


\section{{A radar-holonomic motion out of one of its worldlines} \label{S5}}
As seen in the previous section, the giving of a timelike worldline $z^\mu(t)$ ($t$ is proper time) and the value of the rotation on it, $\Omega_{\mu\nu}(t)$, is not enough to determine a unique radar-holonomic congruence containing that worldline and having that prescribed rotation; rather there is a wide class of radar-holonomic congruences for these data.

In Minkowski spacetime the rotationless Fermi congruence associated to $z^\mu(t)$ ($t$ is worldline's proper time),
\begin{equation} \label{RT1}
\phi^\mu(t,X^i):=z^\mu(t) + X^i\hat{e}^\mu_i(t)  \,,
\end{equation}   
where $\hat{e}^\mu_i$ is a Fermi-Walker transported orthonormal space triad
$$ \frac{d \hat{e}^\mu_i}{dt} = a_i \dot z^\mu \qquad  {\rm with} \qquad a_i:=\ddot z_\nu \hat{e}^\nu_i \,,$$
is a Born-rigid congruence. The radar metric is $\hat g_{ij}=\delta_{ij}$ and the adapted space coordinates $X^i$ are orthogonal.

If the rotation does not vanish, then the transport law is
$$ \frac{d \hat{e}^\mu_i}{dt} = a_i \dot z^\mu + \Omega_i^{\;j} \hat{e}^\mu_j
\qquad  {\rm with} \qquad \Omega_i^{\;j} = \Omega_{ij}= -\Omega_{ji} $$
Written in Fermi coordinates $(\vec X,t)$, Minkowski metric is
\begin{equation} \label{RT2a}
ds^2 = -\xi^2 \,dt^2+ 2  X^l\Omega_{li} \,dt\,dX^i +  \delta_{ij} \,dX^i \,dX^j
\qquad {\rm with} \qquad  \xi^2:= (1+\vec a\cdot\vec X)^2 - X^i X^k \Omega_{ij} \Omega_k^{\;j} \,,
\end{equation} 
and the radar metric (\ref{E.3}) is 
\begin{equation} \label{RT2}
\hat g_{ij} = \delta_{ij} + \xi^{-2} X^lX^k\Omega_{li}\Omega_{kj} 
\end{equation}   
which in general is neither Born-rigid nor radar-holonomic. Indeed, the {\em strain rate}
$$ \Sigma_{ij}:=\partial_t \hat g_{ij} = 2 \xi^{-2} X^l X^k \dot\Omega_{l(i} \Omega_{k|j)} - \frac{2\dot\xi}{\xi^3}\, X^lX^k\Omega_{li}\Omega_{kj} $$
does not vanish unless either ({\em a}) $\Omega_{ij}=0$ or ({\em b}) $\dot a_i=0$ and $\dot\Omega_{ij}=0$. The first case corresponds to a general rotationless motion whereas the second is a Killing motion, in agreement with Herglotz-Noether theorem \cite{HER-NOE10}. However, if the rotation is ``small'', i. e. in the domain range $|\vec X| \ll (\Omega_{ij}\Omega^{ij})^{-1/2}$, the Fermi congruence is approximately Born-rigid, namely, $\Sigma_{ij}= O(\|\Omega\|^2)$.

Although trying to find a perturbation of the Fermi congruence (\ref{RT1}) such that is Born-rigid is nonsense due to  Herglotz-Noether theorem \cite{HER-NOE10}, the search of a perturbation of (\ref{RT1}) which  is radar-holonomic is legitimated by the existence theorem proved in section \ref{S4}.

To set up the perturbative method we shall take that the perturbed congruence moves relatively to the Fermi congruence according to:
\begin{equation}\label{RT5}
\vec X(t,\vec y) = \vec y + \vec P(t,\vec y) \qquad {\rm with} \qquad \|\vec P\| = O(\|\Omega\|^2)
\end{equation}
$\vec y=(y^1, y^2, y^3)=\,$constant corrrespond to a worldline in the perturbed congruence and $(y^1, y^2, y^3, t)$  act as adapted coordinates. We assume the perturbation starts at $O(\|\Omega\|^2)$ because, up to first order, Fermi congruences are Born-rigid. Besides, we shall take $\vec P(t,\vec 0) = 0$ and $\partial_{[i} P_{j]} (t,\vec 0)=0$, so that the origen is not perturbed and the rotation at the origen is $\Omega_{ij}(t)$.

Substituting (\ref{RT5}) into equation (\ref{RT2a}) we have that Minkowski metric written in these adapted coordinates is
\begin{eqnarray*}
ds^2 & =& -\ \left( \xi^2 - 2 \dot{P}^i X^l \Omega_{li} - \dot{\vec P}^2 \right)\,dt^2
+ 2 \,\left( X^l\Omega_{li} + \dot{P}_i \right)\,\left( \delta^i_j + \partial_j P^i\right)\,dt\,dy^j + \nonumber \\
 & &  \left( \delta_{li} +\partial_i P_l \right)\,\left( \delta^l_{j} +\partial_j P^l \right) \,dy^i \,dy^j
\end{eqnarray*}
(latin indices are raised and lowered with $\delta_{ij}$) and, as $\|\vec P\| = O(\|\Omega\|^2)$, we have that
\begin{equation}   \label{RT6}
\widehat{g}_{ij} = \delta_{ij} + 2 \partial_{(i} P_{j)} + \xi^{-2} y^l y^k\Omega_{li}\Omega_{kj} + O(\|\Omega\|^3)
\end{equation}

Now, in order that the perturbed congruence is radar-holonomic, we have to require that three functions $F_i(\vec y,t)$ exist such that $\widehat{g}_{ij}=F_i^2 \delta_{ij}$ and, as the Fermi congruence satisfies this relation up to terms $O(\|\Omega\|^2)$ with $F_i = 1$, for the perturbed congruence we shall take that 
$$ F_i(\vec y,t) = 1 + f_i(\vec y,t) \qquad {\rm with} \qquad f_i = O(\|\Omega\|^2) $$
Substituting this into (\ref{RT6}) we obtain that
\begin{equation}   \label{RT7}
2 \partial_{(i} P_{j)} = 2 f_i \delta_{ij} - \xi^{-2} y^l y^k\Omega_{li}\Omega_{kj} + O(\|\Omega\|^3)
\end{equation}   
which implies that 
\begin{equation}   \label{RT8}
 f_i =  \partial_{i} P_{i} +\frac12\, \xi^{-2} y^l y^k\Omega_{li}\Omega_{ki} + O(\|\Omega\|^3)
\end{equation}
where $P_i$ is a solution of
\begin{equation}   \label{RT9}
2 \partial_{(i} P_{j)} =  - \xi^{-2} y^l y^k\Omega_{li}\Omega_{kj} + O(\|\Omega\|^3)  \,, \qquad  \qquad i\neq j
\end{equation}

After deriving and combining we obtain that these equations imply that
\begin{equation}  \label{RT16b}
\partial_{ij} P_k = \frac12\,\left( \partial_k Q_{ij} - \partial_i Q_{jk} - \partial_j Q_{ki} \right) + O(\|\Omega\|^3) \,, \qquad  i \neq j \neq k 
\end{equation}
where $Q_{ij}:= \xi^{-2} y^l y^r\Omega_{li}\Omega_{rj}$. Notice that these equations do not carry any integrability condition, because only one second derivative of each $P_k$ is known. Writing $\Omega_{ij}=\epsilon_{ijk} \omega^k$ and after a little algebra,we then arrive at
\begin{eqnarray*}
\partial_{ij} P_k &=& \xi^{-3} \left[ y_i y_j \omega_k(2\vec a\cdot\vec\omega - a_k \omega_k) + \left(y_i \omega_j + y_j \omega_i \right) \,\left(\omega_k(1+a_k y_k) - y_k (\vec a\cdot\vec\omega)\right) - \right. \\
 & & \left.\omega_i \omega_j y_k (2 + a_k y_k) \right]  + O(\|\omega\|^3)
\end{eqnarray*}
and on integration we finally obtain:
\begin{eqnarray}  
P_k &=& P_k^{\rm hom} -\frac{(\vec a\cdot\vec\omega) \left[ \omega_k + y_k (\vec a\cdot\vec\omega)\right]}{2a_i a_j} \,M_k +\frac{y_i y_j}{2 \xi \zeta_k \xi_i \xi_j}\,\left\{-\omega_i \omega_j y_k(1 +\zeta_k) (\xi+\zeta_k) + \right. \nonumber \\
& &  \left.  \left[ \omega_k - y_k (a_i \omega_i+a_j \omega_j)\right]\left[ (a_i \omega_i+a_j \omega_j) y_i y_j + \zeta_k(\omega_i y_j + \omega_j y_i)\right] \right\}   + O(\|\omega\|^3)
\label{RT17}  \\[1ex]
{\rm where} & & M_k := \frac{1}{a_i a_j}\,\log\left(1 + \frac{a_i y_i a_j y_j}{\xi\zeta_k}  \right) -\frac{y_i y_j}{\xi\zeta_k}   \,, \qquad i\neq j \neq k \nonumber\\[1ex]
& &  \xi_{i}:=1+\vec a\cdot\vec y-a_i y_i \,, \qquad \zeta_k:=1+a_k y_k  \nonumber
\end{eqnarray}
and $P_k^{\rm hom}=P_{ki}(y_i,y_k,t)+P_{kj}(y_j,y_k,t)$ is a solution of the homogeneous equation which, for the sake of simplicity, we shall choose to vanish.

Then, from (\ref{RT8}) and (\ref{RT17})  we have that
\begin{eqnarray}
f_k & = & -\frac{(\vec a\vec \omega)^2}{2a_ia_j}\left(M_k -\frac{y_iy_j}{\xi\zeta_k}\right) + 
\frac{y_iy_j w_i w_j}{2\xi\xi_i\xi_j\zeta_k}\,\left( [\zeta_k^2-1]\,\left[ \frac{\xi^2+\zeta_k^2}{\xi \zeta_k} + \frac{\xi_i^2+\xi_j^2}{\xi_i \xi_j} \right] - 2[1 +\xi\zeta_k] \right) + \nonumber  \\ & &  \nonumber  \\ 
& \hspace*{-1em}&\frac{y_i^2y_j^2(\xi+\zeta_k)}{2\zeta_k\xi\xi_i\xi_j}\,\left( [\alpha_k\zeta_k - \vec a\cdot \vec\omega]\,\alpha_k\,\left[\frac1{\xi} +\frac1{\xi_i} + \frac1{\xi_j} \right]+\vec a\cdot\vec \omega\,\left[ \frac{\alpha_k -\zeta_k \vec a\cdot\vec \omega}{\xi} - \vec a\cdot\vec \omega \right] \right) +  \nonumber  \\ & & \nonumber  \\
& \hspace*{-1em}& \frac{y_i y_j (w_iy_j+w_jy_i)}{2\xi\xi_i\xi_j}\,\left([\alpha_k\zeta_k - \vec a\cdot \vec\omega]\,\left[\frac1{\xi} +\frac1{\xi_i} + \frac1{\xi_j} \right] -  \alpha_k \right)  + O(\|\omega\|^3) 
\label{final}   
\end{eqnarray}
where $\alpha_k:= \vec a\cdot\vec \omega -a_k\omega_k$.

Finally, the radar-holonomic congruence is
\begin{equation} \label{RT100}
\phi^\mu(t,\vec y):=z^\mu(t) + \sum_{k=1}^3 \hat{e}^\mu_k(t) \left(y_k + P_k(t,\vec y) \right) \,,
\end{equation}   
and the radar metric in coordinates $(\vec y,t)$ is 
$$ \widehat g_{ij}= F_i^2 \delta_{ij}  \qquad {\rm with} \qquad F_i = 1+ f_i(\vec y,t)  $$

\section{Conclusion and outlook}
Looking for a less restrictive substitute for Born's relativistic
definition of rigid motion, we have suggested the definition of {\it
 radar-holonomic } motion, namely, a 3-parameter congruence of timelike
worldlines which admits an adapted system of coordinates
$(t,y^1,y^2,y^3)$, such that the hypersurfaces $y^i=$constant, $i=1,2,3$,
are mutually orthogonal (i.e., the radar metric is diagonal:
$\widehat{g}_{ij}(t,y) = 0$ whenever $i\neq j$).

We have expressed this condition as a partial differential system and
analysed the Cauchy problem. We have proved ---theorem \ref{T6}--- that,
given a 2-parameter congruence ${\cal C}_2$ and a triad of spatial vectors
$\{\overline{\mathbf{e}}_i\}_{i=1,2,3}$ on ${\cal S}_0$ (the track of ${\cal
C}_2$), there is a unique  radar-holonomic  3-congruence, 
${\cal C}_3$, containing ${\cal C}_2$ and admitting an adapted coordinate
system such that the spatial coordinate curves are tangent to
$\overline{\mathbf{e}}_i$ on ${\cal S}_0$.

Moreover, the given directions $\{\overline{\mathbf{e}}_i\}$ are related to the eigenvectors of the strain rate tensor $\overline\Sigma$ and, except in the case that the shear of ${\cal C}_3$ vanishes on ${\cal S}_0$, we have
shown that ${\cal C}_2$ together with $\overline\Sigma$ determine ${\cal C}_3$ in a neigbourhood of ${\cal S}_0$. Hence, a  radar-holonomic congruence is determined by ``a part of it".

The definition of  radar-holonomic  congruences has been devised as an extension to a (3+1)-spacetime of the shear-free congruences studied in reference  \cite{LLOSA97} in a similar context, for the simplified problem of a (2+1)-spacetime. 
We have finally set up a perturbative method to construct congruences of this kind out of one of its worldlines and the value of the rotation on that line.

\section{Acknowledgments}
We are indebted to M. A. Garcia Bonilla for helpful suggestions and comments. JL and AM acknowledge financial support from Ministerio de Educacion y Ciencia through grant no FIS2007-63034 and from the Generalitat de Catalunya, 2009SGR-417 (DURSI).

\end{document}